\documentclass[12pt]{article}

\usepackage[T1]{fontenc}
\usepackage[utf8x]{inputenc}
\usepackage[english]{babel}

\usepackage{graphicx}
\usepackage{epstopdf}

\usepackage{xcolor}

\usepackage{array}
\newcolumntype{P}[1]{>{\centering\arraybackslash}p{#1}}

\usepackage{float}
\usepackage{booktabs}
\usepackage{siunitx}
\usepackage{caption}
\usepackage{subcaption}

\usepackage{amsmath}
\usepackage{amssymb}
\usepackage{amsxtra}
\usepackage{amsfonts}
\usepackage{amsthm}
\usepackage{amsbsy}
\usepackage{upgreek}

\usepackage{mdwlist}
\usepackage{enumerate}
\usepackage{paralist}
\usepackage{multicol}
\usepackage{multirow}
\usepackage{array}
\usepackage{bigdelim}

\usepackage{xspace}

\usepackage[left,mathlines]{lineno}

\def\alphas{\alpha_s}

\def\VVjets{\ensuremath{WZ/ZZ\text{\,+\,jets}}\xspace}

\newcommand\pubnumber{ATL-PHYS-PROC-2021-004}
\newcommand\pubdate{\today}

\def\institute{Fakult{\"a}t f{\"u}r Physik\\
Ludwig-Maximilians-Universit{\"a}t M{\"u}nchen, 85748 Garching, Germany}
\def\support{\footnote{Work supported by BMBF, Germany (FSP-103)}}

\newcommand\blfootnote[1]{%
  \begingroup
  \renewcommand\thefootnote{}\footnote{#1}%
  \addtocounter{footnote}{-1}%
  \endgroup
}
\def\Copyright{\blfootnote{Copyright\ \the\year \ CERN for the benefit of the ATLAS Collaboration. CC-BY-4.0 license.}}

\def\Title#1{\begin{center} {\Large #1 } \end{center}}
\def\Author#1{\begin{center}{ \sc #1} \end{center}}
\def\Address#1{\begin{center}{ \it #1} \end{center}}

\newcommand\pubblock{\rightline{\begin{tabular}{l} \pubnumber\\
         \pubdate  \end{tabular}}}
\newenvironment{Abstract}{\begin{quotation}  }{\end{quotation}}
\newenvironment{Presented}{\begin{quotation} \begin{center} 
             PRESENTED AT\end{center}\bigskip 
      \begin{center}\begin{large}}{\end{large}\end{center} \end{quotation}}
\def\Acknowledgements{\bigskip  \bigskip \begin{center} \begin{large}
             \bf ACKNOWLEDGEMENTS \end{large}\end{center}}

\begin{document}

\begin{titlepage}
\pubblock

\vfill
\Title{Measurement of the inclusive and differential cross section of a top quark pair in association with a $Z$ boson at $13\,\text{TeV}$ with the ATLAS detector}
\vfill
\Author{ Florian Fischer\support, on behalf of the ATLAS Collaboration\Copyright}
\Address{\institute}
\vfill
\begin{Abstract}
The inclusive as well as differential cross section of the associated production of top-antitop quark pairs and a $Z$ boson ($t\overline{t}Z$) is measured in final states with exactly three or four isolated leptons (electrons or muons).
For this purpose, the full LHC Run~2 dataset of proton-proton collisions recorded by the ATLAS detector from \num{2015} to \num{2018}, which corresponds to an integrated luminosity of \SI{139}{\per\femto\barn},  is used.
The inclusive production cross section is measured to be $\sigma_{t\overline{t}Z}=1.05\pm0.05\,(\text{stat.})\,\pm0.09\,(\text{syst.})\,\text{pb}$, which is in agreement with the most precise Standard Model theoretical prediction.
Absolute and normalised differential cross section measurements are performed as a function of various kinematic variables in order to probe the kinematics of the $t\overline{t}Z$ system within both parton- and particle-level phase spaces.
\end{Abstract}

\vfill
\begin{Presented}
$13^\mathrm{th}$ International Workshop on Top Quark Physics\\
Durham, UK (videoconference), 14--18 September, 2020
\end{Presented}
\vfill
\end{titlepage}
\def\thefootnote{\fnsymbol{footnote}}
\setcounter{footnote}{0}

\section{Introduction}
The coupling of the top quark to the $Z$ boson is precisely predicted within the Standard Model (SM) of particle physics by the theory of the electroweak interaction.
However, experimentally it is not yet well constrained and its value can significantly vary in many models including physics beyond the Standard Model (BSM).
A process that is particularly sensitive to this coupling is the associated production of a top-antitop quark pair with a $Z$ boson ($t\overline{t}Z$).
The large centre-of-mass energy of the Large Hadron Collider (LHC)~\cite{lhc} at CERN and the tremendous amount of data collected in recent years have opened up the possibility to study this rare process which was previously inaccessible due to its small production cross section.
As $t\overline{t}Z$ production contributes to the background processes in many searches at the LHC for both SM and BSM physics, a better understanding of the $t\overline{t}Z$ process can further enhance the experimental reach in such analyses.

The results of previous inclusive measurements by the ATLAS~\cite{atlas} and CMS~\cite{cms} collaborations agree very well with the SM prediction~\cite{atlas32,atlas361,cms359}.
A first measurement of differential $t\overline{t}Z$ cross sections was conducted by CMS only recently~\cite{cms775}.
The first analysis using the full LHC Run~2 dataset was performed by ATLAS using \SI{139}{\per\femto\barn} of proton-proton ($pp$) collision data~\cite{atlas139} which is presented in the following.

\section{Analysis channels}
The most sensitive decay channels in which to perform measurements of the $t\overline{t}Z$ process feature a multi-lepton final state with exactly three or four isolated electrons or muons.
Based on these signatures, different signal regions are defined and optimised, referred to as trilepton ($3\ell$) and tetralepton ($4\ell$) signal regions, depending on the respective lepton multiplicity.

Three signal regions are defined for the trilepton decay channel, and four signal regions are defined for the tetralepton decay channel.
Of all lepton pairs with opposite sign of the charge and of the same flavour (OSSF), the one with the value of its invariant mass closest to the $Z$ boson mass is considered to originate from the $Z$ boson decay.
Furthermore, the difference between its invariant mass and the $Z$ boson mass must not be greater than \SI{10}{\giga\electronvolt}.
Contributions from events featuring low-mass resonances are suppressed by requiring all OSSF lepton combinations to have a mass greater than \SI{10}{\giga\electronvolt}.
Additionally, the sum of the lepton charges is required to equal to $\pm1$ and $0$ in the $3\ell$ and in the $4\ell$ case, respectively.
The trilepton signal regions differ from each other by the number of selected jets and $b$-jets, where the latter are tagged with different efficiency working points depending on the required $b$-jet multiplicity.
Similarly, the tetralepton signal regions are categorised into same-flavour and different-flavour regimes of the two non-$Z$ leptons, and each case is again subdivided into a regime with either exactly one or at least two $b$-jets.
In addition, depending on the flavour composition of the non-$Z$ lepton pair and $b$-jet multiplicity, different thresholds on the missing transverse energy are required.

\section{Background estimation}
Background processes -- physics processes described by the Standard Model other than $t\overline{t}Z$ -- are subdivided into prompt and non-prompt contributions.

The dominant prompt background processes are \VVjets production which feature either three or four isolated leptons in the final state, respectively.
Dedicated control regions are used to estimate the light-flavour components of these backgrounds during the fit employed for the inclusive cross-section measurement.
These regions are defined such that they are orthogonal to the respective signal regions and are predominantly populated with events featuring \VVjets light-flavour components.
In contrast, the charm- and bottom-flavour components are constrained in the fit with the corresponding uncertainties assigned which are related to the simulation of heavy-flavour components.
Further SM background processes considered such as the associated production of single top quarks or top-antitop quark pairs with heavy vector bosons are estimated directly from simulated Monte Carlo (MC) samples.

Background contributions from leptons from secondary decays (``non-prompt'') or so-called fake leptons (objects misidentified as leptons), however, are estimated employing a data-driven method, referred to as matrix method.
Details about this method can be found in the reference documents~\cite{matrix1} and~\cite{matrix2}.

\section{Results}
The inclusive $t\overline{t}Z$ production cross section is extracted by performing a simultaneous maximum-likelihood fit to the number of events in the trilepton and tetralepton signal regions, as well as the \VVjets control regions.
A total of three free parameters are given to the fit: the ratio between the measured value of in the inclusive $t\overline{t}Z$ production cross section and its corresponding Standard Model prediction, referred to as signal strength, as well as the normalisation factors of the \VVjets backgrounds used to extrapolate the corresponding event yields into the signal regions.
The inclusive $3\ell+4\ell$ cross section of $t\overline{t}Z$ production in $pp$-collision data at a centre-of-mass energy of \SI{13}{\tera\electronvolt} is measured to be:
\begin{equation}
\sigma(pp\to t\overline{t}Z)=1.05\pm0.05\,(\text{stat.})\,\pm0.09\,(\text{syst.})\,\text{pb}=\left(1.05\pm0.10\right)\,\text{pb}
\end{equation}
This result agrees with the dedicated theoretical prediction~\cite{theo-incl} of
\begin{equation}
\sigma_{t\overline{t}Z}^{\mathrm{NLO+NNLL}}=0.863^{+0.07}_{-0.09}\,(\mathrm{scale}) \pm 0.03 \,(\mathrm{PDF+\alphas})\,\mathrm{pb}\quad.
\end{equation}
The uncertainties on this result are dominated by the systematic uncertainties of which the most important ones are related to the modelling of the parton shower in the signal Monte Carlo, the modelling of various background processes, and the $b$-tagging procedure.

In addition to the inclusive result, the $t\overline{t}Z$ cross section is measured as a function of different variables sensitive to the kinematics and the production of the $t\overline{t}Z$ system.
For this purpose, a total of nine such variables are unfolded to parton and particle level, employing the Iterative Bayesian Unfolding method~\cite{ibu}.
On parton level, the (anti-)top quark and the $Z$ boson can be directly accessed before the decay within Monte Carlo simulation, whereas on particle level these have to be reconstructed from simulated stable particles without any modelling of their interaction with the detector material or pile-up.
In this way, events are corrected for detector effects and results can be directly compared to theoretical calculations.
\begin{figure}[!htb]
\subfloat[]{\includegraphics[width=0.49\textwidth]{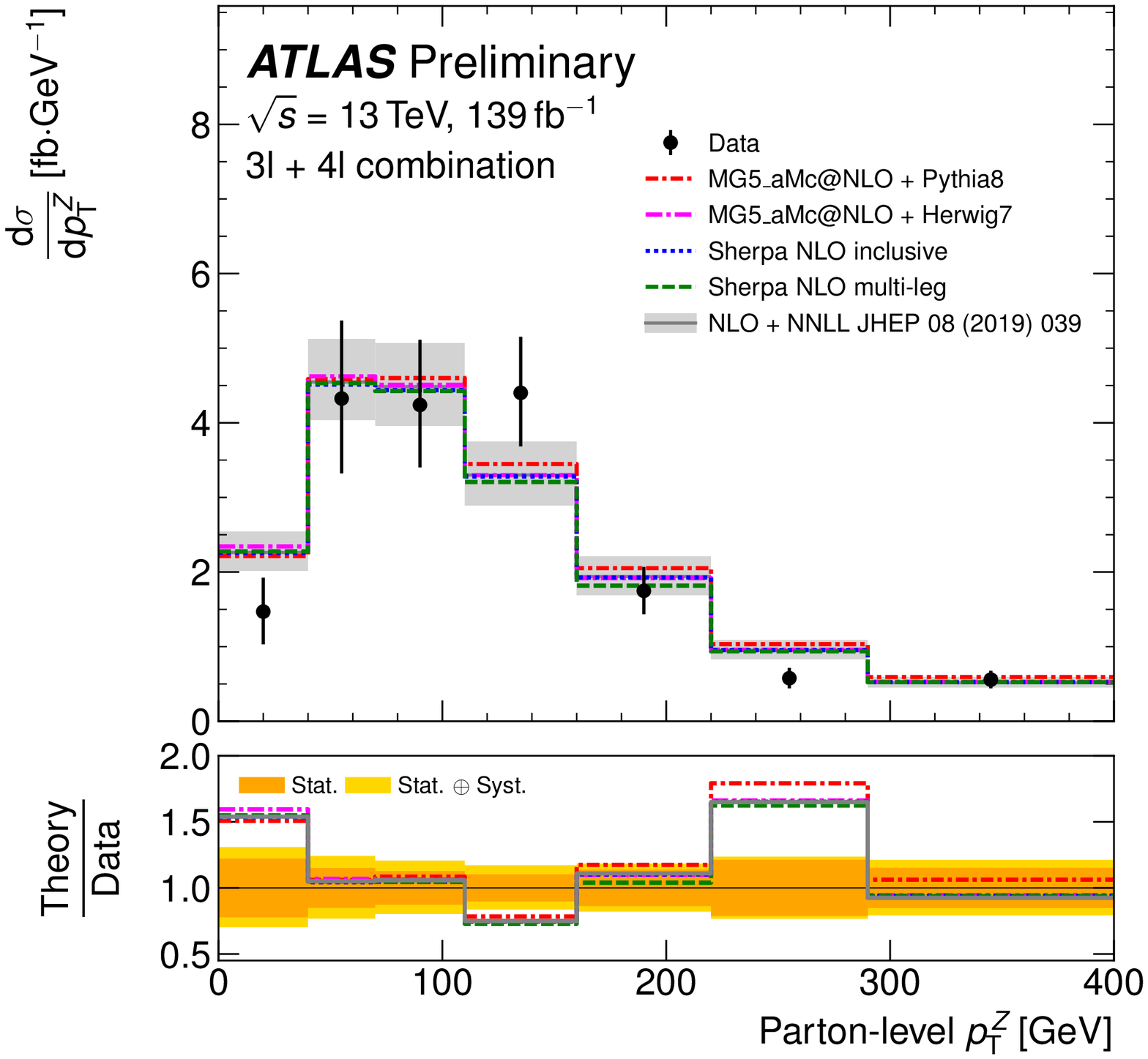}  \label{fig:diff-sec-abs}}
\subfloat[]{\includegraphics[width=0.49\textwidth]{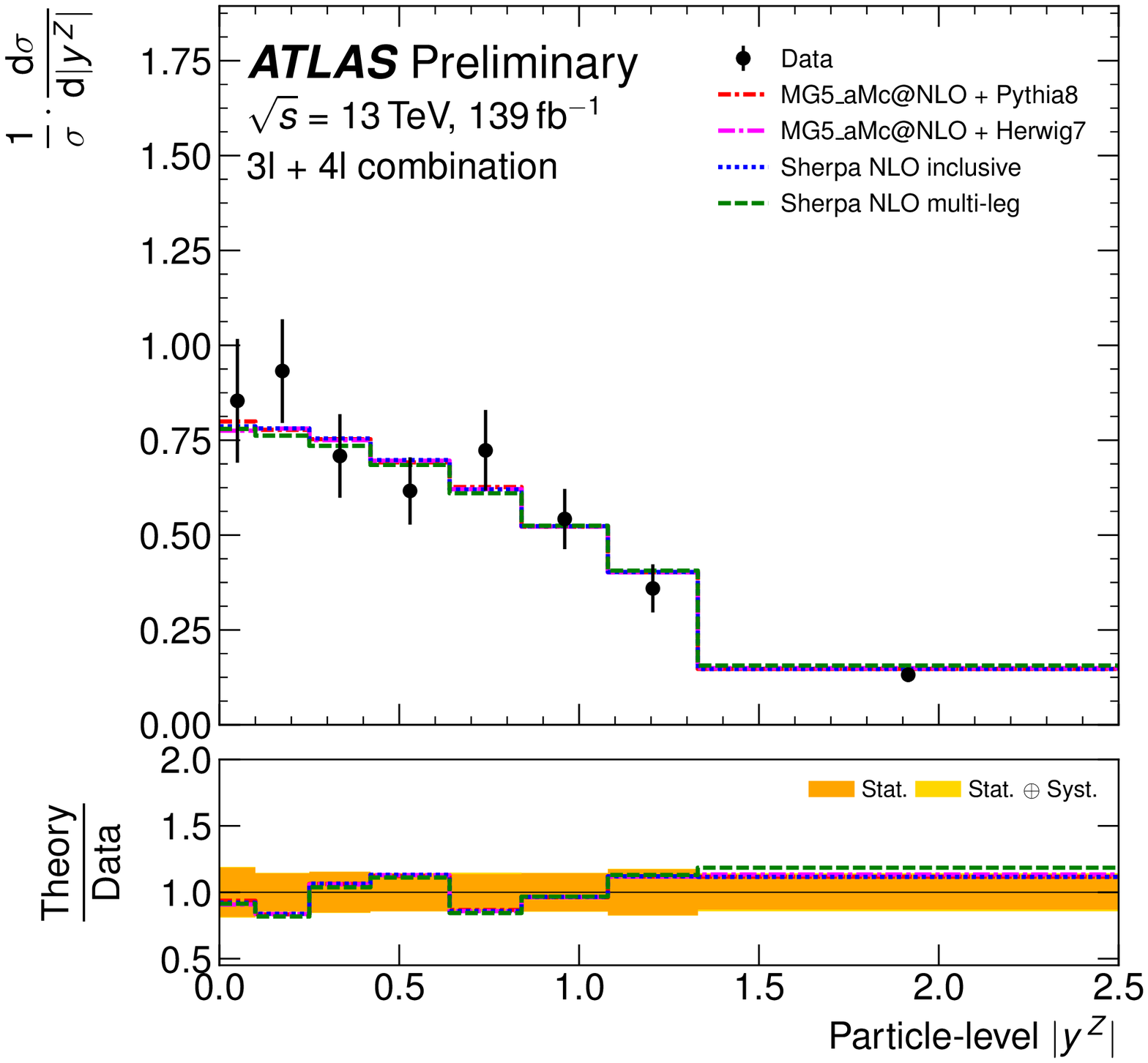} \label{fig:diff-sec-norm}}
\caption{Absolute (left) and normalised (right) cross section measured at parton (left) and particle (right) level as a function of the transverse momentum (left) and of the absolute rapidity (right) of the $Z$ boson.
The predictions of various MC generators are represented by dashed and dotted coloured lines whereas the data are depicted as black dots.
In addition, custom differential~\cite{theo-diff} predictions are shown by a black solid line within a grey-shaded area.
In the ratio panels, the relative contributions from both the statistical and systematic uncertainties are shown.
A branching fraction of $\mathcal{B}(t\overline{t}Z_{3\ell+4\ell})=0.0223$ is applied for the parton-level result~\cite{atlas139}.}
\label{fig:diff-xsec}
\end{figure}
This analysis determined both the absolute and normalised differential cross section for the $3\ell$ and $4\ell$ scenarios separately as well as for the combination.
In Figure~\ref{fig:diff-xsec}, two examples for the differential cross section measurements in the combined $3\ell+4\ell$ channel are depicted.
Different simulated samples generated with different sets of MC generators, including the nominal \textsc{MG5\_aMC@NLO+Pythia~8}~\cite{amcatnlo,pythia},  as well as a set of additional differential predictions, calculated at parton level as described in~\cite{theo-diff}, are compared to the unfolded data.
In general, a good agreement between the unfolded data and the various predictions can be observed.

\Acknowledgements
The author would like to thank for the support of his work by BMBF, Germany (FSP-103).

\end{document}